\documentclass[12pt]{article}

\usepackage[margin=1.4cm]{geometry}
\usepackage{multirow}
\usepackage{enumitem}
\usepackage{amssymb}
\usepackage{amsthm}
\usepackage{amsmath}
\usepackage{graphicx}
\usepackage{array} 
\usepackage{subcaption}
\usepackage{lipsum} 
\usepackage{setspace}
\usepackage[sort&compress,numbers]{natbib}
\usepackage[colorlinks=true, linkcolor=blue, citecolor=blue, urlcolor=blue]{hyperref} 

\usepackage{multirow}
\usepackage{booktabs}
\usepackage{tabularx}
\graphicspath{ {figures/} }
\usepackage{float}

\usepackage{tikz}
\usetikzlibrary{shapes.geometric, arrows.meta, positioning}
\usepackage{algorithm}
\usepackage{algorithmic}
\usepackage{pdflscape}
\usepackage{graphicx}

\usepackage{tikz}
\usetikzlibrary{shapes.geometric, arrows, positioning}

\tikzstyle{process} = [rectangle, minimum width=3cm, minimum height=1cm, text centered, draw=black]

\title{Tabu Search for Tactical Wireless Network Design in Challenging Environments}
\author{
	Wissem Ahmed Zaid,
	Alain Hertz
	\\[3mm]
	   \footnotesize Department of Mathematics and Industrial
	Engineering\\[-2mm]
	\footnotesize Polytechnique Montr\'eal - Gerad, Montr\'eal, Canada\\[-2mm]
	\footnotesize Corresponding author: alain.hertz@gerad.ca\\[3mm]
}

\date{\today}

\begin{document}

\maketitle

\begin{abstract}
Tactical wireless networks play a vital role in ensuring reliable connectivity in scenarios where conventional telecommunications infrastructure is unavailable or damaged, such as areas impacted by natural disasters.
These networks are designed to operate efficiently in difficult and unpredictable environments by adapting to the unique characteristics of the terrain.
This research addresses a real-world challenge from the communications industry: designing tactical wireless networks that meet the specific constraints defined by our industrial partner, with the goal of optimizing signal strength and coverage while minimizing interference. To this end, we propose two tabu search algorithms that incorporate several heuristic subroutines, enabling the efficient generation of high-quality network designs. Results from synthetic tests demonstrate that our approach produces networks rapidly and effectively, offering significant improvements over existing methods.

\vspace{0.3cm}\noindent\emph{Keywords:} network design; tactical wireless networks; tree topology; tabu search.

\end{abstract}

\section{Introduction}
Wireless communication is central to modern information technology, typically relying on conventional telecommunication networks. When these networks are unavailable or disrupted, such as during natural disasters, establishing temporary tactical networks becomes essential to maintain communication. The goal is to connect key locations, represented as network nodes, to ensure reliable data exchange.

Our research focuses on designing tactical wireless networks that meet the specific requirements of a real-world industrial problem in the communications sector.
Such networks typically link 10 to 50 nodes across a given region. Each node is equipped with a radio and a pair of multi-beam antennas. Each radio has two channels (one per antenna) and supports two signal frequencies per channel. 
The network is modeled as a directed, rooted tree in which all edges are oriented away from the root node. This root node is designated as the master hub, serving as the central point of control and coordination within the network.
Communications between nodes can be carried out in two different ways:
Point-to-Point (PTP) allows establishing direct connections between two nodes, providing direct and reliable communication over long distances; Point-to-MultiPoint (PMP) allows a node to communicate with multiple remote nodes simultaneously, thus enabling efficient data distribution over a wide geographic area.
The network’s bottleneck is defined by its weakest connection in terms of effective throughput. The goal is to maximize this bottleneck to achieve the highest possible worst-case data transmission rate, ensuring that critical information can be transmitted between all nodes as efficiently as possible.

In tactical scenarios like disaster relief operations, it is essential to account for the terrain's specific physical characteristics when determining the most effective network topology. It is, for example, important to consider signal loss during communications, as it may occur due to factors such as interference. Most existing studies tend to overlook these factors and rely on simplifying assumptions to enable the use of standard solution methods, such as integer linear programming. However, as noted in \cite{2}, the problem becomes highly nonlinear and challenging to solve without these simplifications. The aim of this paper is to present an algorithm capable of quickly providing good solutions to this complex problem.

The paper is organized as follows. Section~\ref{sec:3} provides a detailed description of the network design problem under consideration. Section~\ref{sec:2} offers a brief review of the related literature, highlighting the simplifying assumptions made in previous studies. Section~\ref{sec:4} presents the problem formulation and clearly identifies the factors contributing to its complexity. In Section~\ref{sec:6}, two algorithms are proposed and subsequently compared with each other and with existing methods in Section~\ref{sec:7}, using synthetic data. Finally, Section~\ref{sec:8} concludes the paper with key remarks and outlines directions for future research.

\section{Problem description}
\label{sec:3}

An instance of the problem is specified by a set 
$V$ of nodes, where each node $v\in V$ has known coordinates.
 The network we aim to design must have a tree structure. After forming a tree that connects all nodes in 
$V$, one node is chosen as the master hub, and the tree is oriented by directing all edges away from this central node.
Each node includes a radio linked to two multi-beam antennas. Each radio is equipped with two channels, each linked to a separate antenna and capable of operating on two frequencies. The master hub’s successors are split into two groups, each assigned to one of its channels.
Each of the remaining nodes has exactly one direct predecessor and may have multiple successors. One channel is dedicated to the connection with its predecessor, and the other is allocated to the connections with its successors.
When there is only one successor, the link operates in point-to-point (PTP) mode; it operates in point-to-multipoint (PMP) mode when there are two or more successors.
Next, the antennas are configured by minimizing the number of active beams, since adding more beams weakens the signal strength for all connected edges.

An additional constraint must be taken into account. Due to technological limitations, the radio interface of a node, when operating on one of the two available channels, can communicate with at most ten other nodes. As a result, the degree of the master hub is limited to 20, whereas the degree of any other node cannot exceed 11, one link corresponding to the channel used for communication with its predecessor in the rooted directed tree, and up to ten links corresponding to the channel used for communication with its successors.

\begin{figure}[!htb]	\centering    \includegraphics[scale=0.45]{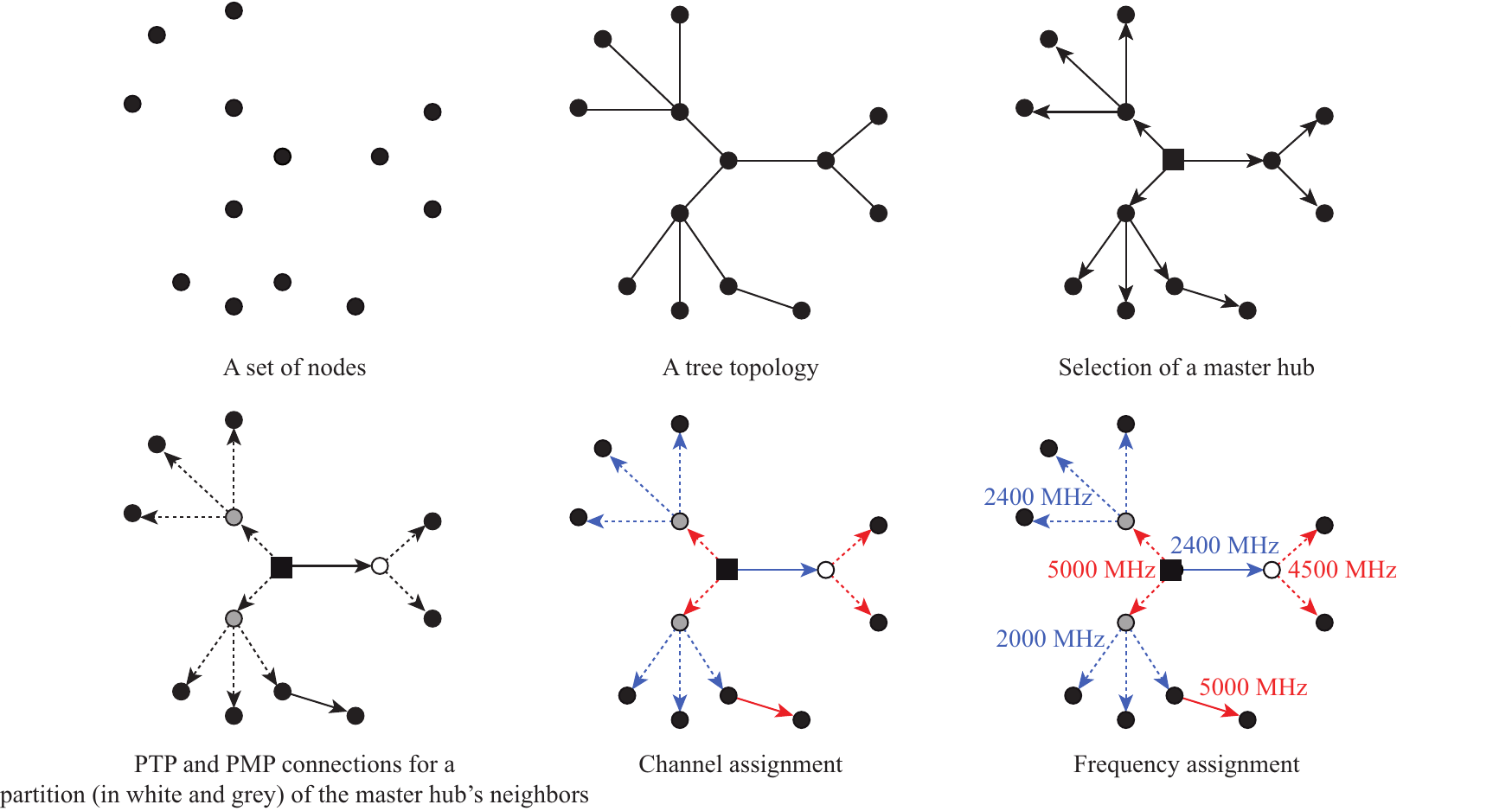}
\caption{Illustration of the network design process.}
	\label{fig:process}
\end{figure}

Figure~\ref{fig:process} illustrates the network design process.
Beginning with a set of nodes, a tree topology 
$T$ is first constructed. A master hub 
$r$ is then selected, shown as a black square.
 The three neighbors of the master hub are then divided into two sets: one containing the two grey nodes, and the other containing the white node. Solid lines indicate PTP connections, while dashed lines indicate PMP connections.
Then channels are assigned, with one channel represented in red and the other in blue.
Finally, frequencies are assigned: assuming that the red channel operates at 4500 MHz and 5000 MHz, while the blue channel operates at 2000 MHz and 2400 MHz. The activated antenna beams and their alignments are omitted in this Figure, since a straightforward geometry-based procedure \cite{2} configures each antenna for connections between  a node to its direct successors.

The direct throughput 
$TP_{uv}$
 can then be computed for each edge $uv$ in the tree topology. Reference \cite{2} presents a detailed procedure for calculating these throughputs by modeling the physical signal and incorporating the interference effects between edge pairs that share the same frequency.
Three traffic modeling scenarios are considered. Let 
$n^X_{uv}$ denote the number of data streams in scenario 
$X$ for an edge 
$uv$ of the tree, and let 
$d_v$ be the number of nodes 
$w$ for which a path exists from 
$v$ to 
$w$ in the rooted directed tree.
\begin{itemize}\itemsep=-1pt
\item Scenario A: There is a single data stream between any two nodes in the network (only one stream is active at a time). Consequently, 
$n^A_{uv}=1$ for all edges 
$uv$ in the tree.
\item Scenario B: There is a data stream from the master hub to every other node ($|V|-1$ streams active simultaneously). For each edge 
$uv$, 
$n^B_{uv}=d_v$ if the edge is directed from 
$u$ to 
$v$, and 
$n^B_{uv}=d_u$ otherwise.
\item Scenario C: There is a data stream from every node to every other node (
$|V|(|V|-1)$ streams active simultaneously). Each edge carries all signals from its descendants to the rest of the nodes, as well as signals coming from the opposite direction. Hence, 
$n^C_{uv}=2d_v(|V|-d_v)$ if the edge is directed from 
$u$ to 
$v$ , and 
$n^C_{uv}=2d_u(|V|-d_u)$ otherwise.
\end{itemize}
 
The goal is to maximize the minimum effective throughput, thereby avoiding bottlenecks, as well as the average throughput, to achieve balanced global performance.
The objective function depends on the tree topology 
$T$, the chosen master hub 
$r$, the partition 
$\pi$ of 
the master hub's successors, the channel assignment 
$\sigma$, the frequency assignment 
$\varphi$, and the antenna configurations 
$\alpha$.
By weighting the relative importance of the average throughput versus the minimum throughput through a parameter $p$, and by associating a weight $\omega_X$ with each scenario 
$X\in\{A,B,C\}$, the resulting function for optimization is:
$$O(T,r,\pi,\sigma,\varphi,\alpha)=\sum_{X\in\{A,B,C\} } \omega_X \left(
\min_{[u,v] \in E} \frac{TP_{uv}}{n^X_{uv}}
+ 
p\;\text{mean}_{[u,v] \in E} \frac{TP_{uv}}{n^X_{uv}}
\right)
$$
where the throughputs $TP_{uv}$ depend on $T,r,\sigma,\varphi$ and $\alpha$,
the weights $\omega_X$ reflect the relative importance of each scenario, and $p$ controls the relative importance of the average throughput versus the minimum throughput. For the calculation of $TP_{uv}$, the reader is referred to the Appendix (Section \ref{appendix}).

\section{Literature review}
\label{sec:2}

The design of telecommunication networks spans a broad range of applications, many of which differ considerably from the tactical context addressed in this work. For instance, some studies focus on static or long-term communication infrastructures, such as networks based on free-space optical links \cite{8,9}, fiber optics \cite{10,19}, or the LoRa protocol \cite{11}. These approaches typically emphasize factors like installation cost, energy consumption, reliability, and overall network capacity.
Other research targets radio-based network design with distinct structural paradigms, including multi-tree configurations \cite{12,13}, star-of-stars topologies \cite{14}, or cluster- and routing-based architectures \cite{15}. Such studies generally aim to minimize infrastructure and energy costs, sometimes at the expense of overall network performance.
Further works explore optimization-based methods for wireless network design, such as reducing installation costs and interference via integer programming and heuristics \cite{15KEDANE}, applying multi-criteria decision techniques to balance cost, reliability, and throughput \cite{21KEDANE}, or employing topology control and routing strategies to enhance capacity, scalability, and energy efficiency in multihop and sensor networks \cite{22KEDANE,23KEDANE,26KEDANE,27KEDANE}.
Beyond these domains, network design principles have also been examined in other telecommunication-related systems, including fleet coordination between ships and control vessels \cite{16}, swarm formation management for mobile robots \cite{17}, and information exchange among autonomous communicating agents \cite{18}. Each application presents unique structural and operational challenges, underscoring the diversity and inherent complexity of telecommunication network design.

As emphasized by our industrial partner, in realistic deployment scenarios it is crucial to explicitly account for antenna orientation, beam patterns, PTP and PMP connections, and frequency-based interference management. Most existing studies on network design tend to overlook one or more of these important aspects.
For example, the authors in \cite{hamami, zhou, zhourooting} model interference simply as a binary condition, classifying links as either conflicting or non-conflicting. A refined approach is proposed in \cite{marina} using a conflict graph, where each link is assigned a conflict weight aggregating interference from all other conflicting links; the optimization then focuses on minimizing the maximum conflict weight across channels rather than maximizing throughput. Also, while \cite{mumey}  uses a physically based signal model, it assumes idealized conditions with no inter-link interference or terrain-induced signal loss. As another example of a neglected constraint, the model in \cite{zhourooting} relies on single-beam antennas with fixed orientations and address interference solely by assigning different channels (while only two are available in a realistic setting). Also, \cite{hamami} and \cite{zhou} consider single-beam antennas with channels but no frequencies, while \cite{mumey} allows multi-beam antennas yet ignores both channels and frequencies. 
A method for designing networks with a tree topology, connecting a base station controller to multiple transceiver stations, is proposed in \cite{li}: the aim is to minimize installation costs and communication delays, considering only omnidirectional antennas and a simplified binary signal model. In a same spirit, \cite{huang} investigates maximizing network capacity in an acyclic topology, under the same antenna and signal model assumptions.
Although \cite{huang2} aims to maximize throughput in wireless sensor networks, it overlooks the realistic impact of interference, which may significantly affect performance.

Exact MIP-based methods have been proposed, for example, in \cite{hamami, zhou, mumey, zhourooting, kedane}, but they simplify the problem by modeling antenna alignments as a limited number of idealized sectors (4–12), assuming uniform radiation within each sector. This simplification neglects that real antennas focus most energy along the beam’s central axis, with power tapering toward the edges.
In our work, we consider the full range of alignment angles [0,2$\pi$] and employ a realistic radiation pattern where signal strength peaks at the beam axis and decreases with angular deviation. This allows the optimization to capture alignment precision and true signal variability, yielding solutions that more closely reflect practical deployment.


To our knowledge, the only work that fully addresses this problem without overlooking any constraints is Vincent Perreault's master thesis \cite{2}. In it, he proposes an exact algorithm capable of solving problems with up to 10 nodes, as well as a parallel tabu search method to generate solutions for larger instances. This paper aims to introduce a heuristic that improves upon Perreault’s approach.

\section{Algorithmic tools}
\label{sec:4}

In this section, several fundamental tools that will be employed in the algorithms presented in Section 5 are introduced. The discussion begins with a number of definitions that serve to clarify the operations permitted at each step of the process illustrated in Figure \ref{fig:process}.
 \begin{itemize}\itemsep=-1pt
\item \emph{valid topology}: A valid topology is a tree that connects all given points. At most one vertex may have a degree greater than 11, and if so, its degree must not exceed 20.

    \item \emph{valid master hub}: A valid master hub of a valid topology is either the unique vertex with a degree greater than 11, if such a vertex exists, or any vertex otherwise.

    \item \emph{valid partition}: A partition of the master hub’s neighbors is considered valid if it is composed of two groups, each with no more than 10 nodes.

    \item \emph{valid channel assignment}: A channel assignment is said to be valid if, considering the directed tree rooted at the master hub $r$, the unique incoming arc to any point other than $r$ uses one of the two available channels, while the other channel is used for connections to that point’s successors. For the master hub, each group of a valid partition of its neighbors is assigned one of the two channels.

    \item \emph{valid frequency assignment}: A frequency assignment is considered valid if every edge of the tree is assigned one of the two frequencies allocated to its corresponding channel.

\end{itemize}

\begin{figure}[!htb]
	\centering    \includegraphics[scale=0.9]{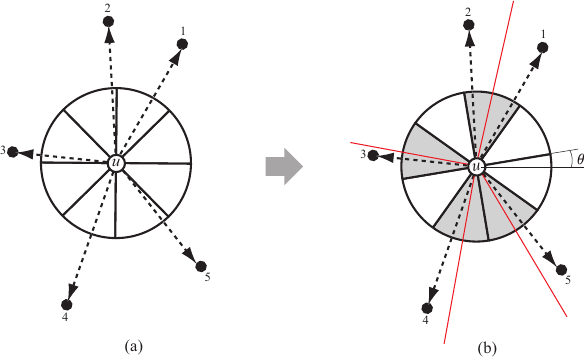}
\caption{Example of an antenna configuration.}
	\label{fig:antennatconfiguration}
\end{figure}

Given a node $u$ and its set of successors with which $u$ must communicate over a given channel, there exists a procedure described in \cite{2} that allows configuring the antenna at $u$ used to enable these communications. This procedure defines the antenna alignment and determines which beams are active. For illustration, Figure \ref{fig:antennatconfiguration}(a) shows a node 
$u$ with 5 successors, before configuring its 8-beam antenna. The result of the procedure described in \cite{2} is shown in Figure \ref{fig:antennatconfiguration}(b): the antenna has been rotated by an angle 
$\theta$, and 4 beams are activated. The red lines indicate the directions of the emitted signals.

Various factors contribute to the complexity of the problem we aim to solve. First of all, let 
$f(T)$ denote the maximum value of 
$O(T,r,\pi,\sigma,\varphi,\alpha)$ over all valid choices of $r,\pi,\sigma,\varphi,\alpha$. The problem to be addressed, therefore, consists of identifying a valid topology 
$T$ that maximizes the value of 
$f(T)$. The difficulty arises from the fact that, for any given topology 
$T$, determining the corresponding value of 
$f(T)$ is far from straightforward. This complexity stems not only from the large number of possible choices for the parameters 
$r,\pi, \sigma, \varphi,\alpha$, but also from the fact that evaluating the function 
$O(T,r,\pi,\sigma,\varphi,\alpha)$ requires computing the throughput $TP_{uv}$ on each edge $uv$ of $T$, a task that is computationally intensive and by no means trivial. As noted earlier, the detailed procedure for calculating a throughput is provided in \cite{2}, and we also include this computation in the Appendix (Section \ref{appendix}) for reference. Given the inherent complexity of this evaluation, and for the sake of clarity and focus in the remainder of this article, we adopt the simplifying assumption that, once the topology 
$T$ and the associated parameters 
$r,\pi, \sigma, \varphi,\alpha$ are specified, the value of 
$O(T,r,\pi,\sigma,\varphi,\alpha)$ can be obtained directly from a “black box.” This abstraction allows us to concentrate on the combinatorial and optimization aspects of the problem without being encumbered by the detailed and computationally heavy throughput calculations at every step.

The other factors that play a particularly significant role in contributing to the overall complexity of the problem can be identified as follows:
\begin{itemize}\itemsep=-1pt
    \item The number of trees that can be built on 
$n$ vertices is 
$n^{n-2}$, and most of them represent valid topologies.
\item The number of valid partitions of the master hub’s neighbors grows exponentially with its degree. For instance, there are 
$\frac{1}{2}\binom{20}{10}=92,378$ valid partitions when the degree is 20 (yielding two blocks of 10 neighbors), and 
$68,068$ valid partitions when the degree is 18 (with blocks of either 9–9 or 8–10 neighbors).
\item Given a valid topology with a valid master hub, a valid partition of its neighbors, and a valid channel assignment, there exists an exponential number of valid frequency assignments, since each channel admits two possible frequency choices. Thus, for example, if the tree is a chain on 
$n$ vertices, the number of valid frequency assignments is 
$2^{n-1}$.
\end{itemize}

Several procedures are introduced in \cite{2} to overcome the challenges arising from this complexity. An initial approach consists in reducing the number of pairs of nodes eligible to be connected by an edge in a valid topology. More specifically, for any two nodes 
$u$ and $v$, one can compute the throughput 
$TP_{uv}$ under the assumptions that the antenna involved in the communication between these two nodes is perfectly aligned, that the highest frequency is used, and that no interference affects the link. If the resulting throughput 
is zero, then the pair 
$(u,v)$ is included in the set 
$E_{not}$, which contains all pairs of nodes for which establishing an edge in a valid topology is not meaningful.

The following method, as suggested in \cite{2}, allows to avoid examining the exponential number of possible frequency assignments. Given a valid topology $T$ with a valid selected master hub $r$, a valid partition $\pi$ of its neighbors, a valid channel assignment $\sigma$, and an antenna configuration $\alpha$, the original objective function can be replaced with a coarse estimate in which the throughput on each edge of the topology is computed by ignoring interference and taking the average throughput over the two possible frequencies. This avoids having to consider the $2^{n-1}$ possible frequency assignments. We thus obtain an estimate of 
$\max\limits_{\mbox{\scriptsize valid }\varphi}O(T,r,\pi,\sigma,\varphi,\alpha)$, which we denote 
$O^{Est}(T,r,\pi,\sigma,\alpha)$.

As described in \cite{2}, it is also possible to compute a lower bound on the original objective function using a greedy algorithm for the frequency assignment. Specifically, given 
$T$, $r$, $\pi$, $\sigma$, and $\alpha$, a valid frequency assignment 
$\varphi$ is constructed by processing the connections in a breadth-first order \cite{thomas2009introduction}, starting at the master hub and progressing systematically toward the leaves of the topology. At each step of the process, for the connection under consideration, we evaluate all valid frequency options and select the one that yields the highest value of the objective function for the subtree consisting only of edges with assigned frequencies. In this way, we progressively construct a complete assignment that provides a meaningful lower bound on the overall performance measure. This approach once again removes the necessity of exhaustively evaluating all 
$2^{n-1} $possible frequency assignments. By bypassing this combinatorial explosion, the method is able to produce a meaningful and computationally efficient lower bound on 
$\max\limits_{\mbox{\scriptsize valid }\varphi}O(T,r,\pi,\sigma,\varphi,\alpha)$, which we denote 
$O^{LB}(T,r,\pi,\sigma,\alpha)$.

\begin{figure}[!htb]
	\centering    \includegraphics[scale=0.50]{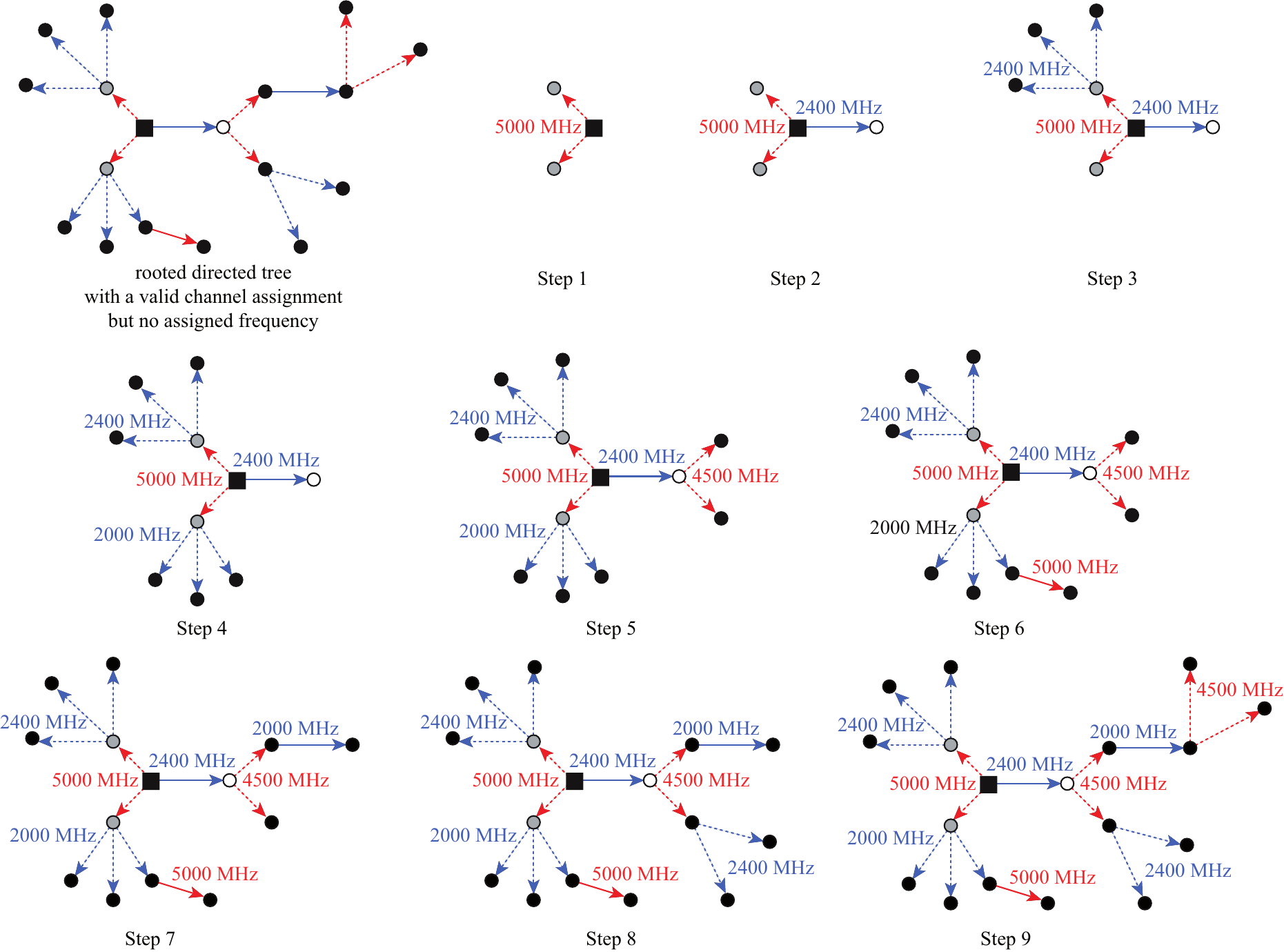}
\caption{Illustration of the greedy algorithm for the frequency assignment.}
	\label{fig:frequencygreedy}
\end{figure}

The greedy algorithm is illustrated in Figure \ref{fig:frequencygreedy}. In the upper-left portion of the figure, the network topology is displayed, showing the master hub (represented by the black square), together with one of its partitions of neighboring nodes (indicated by the white and grey circles). Each edge connecting the hub to its neighbors is shown with its assigned channel, illustrated using red and blue colors. The red channel can take frequencies of 4500 MHz or 5000 MHz, and the blue channel 2000 MHz or 2400 MHz. The algorithm first selects 5000 MHz for the red links from the master hub to the two nodes on its left, then chooses 2400 MHz for the blue link to its right. The procedure then continues in the same manner through each successive step, eventually reaching step 9, at which point every connection in the system has been assigned a frequency.

\section{Three tabu search algorithms}
\label{sec:6}

In this section, three tabu search algorithms are described. The first is the one proposed in \cite{2}, which motivated the development of the other two presented in this paper, as the original algorithm could perform too few iterations within a reasonable amount of time.

\subsection{A Tabu Beam Search}\label{sec:tbs}
The algorithm proposed in \cite{2} is a form of parallel tabu search, known as Tabu Beam Search (\texttt{TBS}), which combines the principles of tabu search with a beam search strategy to explore multiple solution paths simultaneously.

The solution space explored by \texttt{TBS} consists of all valid topologies that do not contain any edges from the set 
$E_{not}$. Given a valid topology 
$T$, only a subset 
$R(T)$ of its nodes can be chosen as the master hub. If 
$T$ contains a vertex 
$v$ with degree greater than 11, then 
$R(T)=\{v\}$. Otherwise, let 
$W$ be the set of non-leaf nodes in 
$T$. For each 
$v\in W$, define 
$n(v)=p_d(v)+p_e(v)$
, where 
$p_d(v)$ is the node’s position in 
$W$ ordered by increasing degree, and 
$p_e(v)$ its position in 
$W$ ordered by decreasing eccentricity. Nodes with high 
$n(v)$ have both high degree and central location. 
$R(T)$ is then defined as the set of nodes 
$v\in W$ with 
$n(v)\geq \textrm{median}_{u\in W}\,n(u)$.

A neighbor of a given topology is created by deleting one of its edges and adding a different edge that maintains the connectivity of the graph. Only those resulting neighbors that yield valid topologies are taken into account. Two tabu lists are used: one forbidding the removal of an edge, and the other forbidding the addition of an edge. When a neighbor $T'$ of $T$ is selected for the next iteration, the edge that was removed is placed on the list forbidding its addition, while the edge that was inserted is placed on the list forbidding its removal.

As mentioned in Section \ref{sec:4}, 
$f(T)$ is the maximum value of 
$O(T,r,\pi,\sigma,\varphi,\alpha)$ over all valid choices of $r,\pi,\sigma,\varphi,\alpha$. To estimate the value $f(T')$ of a neighbor topology 
$T'$ of $T$, each vertex 
$r\in R(T')$ is considered in turn as the master hub. All valid partitions 
$\pi$ of the neighbors of $r$ are tested, the antennas are configured accordingly, and both channel assignments 
$\sigma$ are examined. The estimated value $f(T')$ of 
$T'$, denoted $f^{Est}(T')$,
 is then taken as the maximum of 
$O^{Est}(T',r,\pi,\sigma,\alpha)$ over all tested 
$r$, $\pi$, and $\sigma$.

At each iteration, \texttt{TBS} maintains a set 
$\mathcal{T}=\{T_1,T_2,\ldots,T_{\kappa}\}$
 of 
$\kappa$ topologies. For each 
$T_i$ in this set, the algorithm generates all valid neighboring topologies and estimates their values using the function 
$f^{Est}$ as described above. Only the 
$\kappa$ neighbors $T'$ of 
$T_i$  with the highest values $f^{Est}(T')$ are retained. In total, this produces 
$\kappa^2$ neighboring topologies ($\kappa$ neighbors for each topology in 
$\mathcal{T})$. From these 
$\kappa^2$ topologies, only the 
$\kappa$ topologies with the highest 
$f^{Est}$ values are kept for the next iteration, forming the new set 
$\mathcal{T}$.

A lower bound $f^{LB}(T_i)$ on the value $f(T_i)$ of each topology $T_i$ in 
$\mathcal{T}$
 is obtained using the function 
$O^{LB}$. More precisely, the estimated value 
$f^{Est}(T_i)$ of each 
$T_i\in \mathcal{T}$ was obtained with a specific choice of master hub 
$r$
and a partition 
$\pi$ of its neighbors, which induced an antenna configuration 
$\alpha$. For the same choice of 
$r$ and 
$\pi$, both possible channel assignments 
$\sigma$ are tested, and the greedy algorithm described in the previous section assigns frequencies 
 to obtain a lower bound 
$O^{LB}(T_i,r,\pi,\sigma,\alpha)$
on $f(T_i)$. If the master hub 
$r$ has at most 7 neighbors, all other partitions of its neighbors are also tested; otherwise, only a few alternative partitions obtained from 
$\pi$ by moving one neighbor of $r$ from one block of the partition to another are considered. Among all these lower bounds, the highest value constitutes the best bound and is denoted 
$f^{LB}(T_i)$. 

For every frequency 
$\mathrm{f}$ and every pair of nodes 
$u$ and $v$, the quantity 
$\ell_{uv}^\mathrm{f}$ specifies how much signal is lost during communication between the two nodes at that frequency.
Considering the set 
$F$ of the four available frequencies (two per channel), and defining 
$\Bar{\ell}_{uv}$ as the average of 
$\ell_{uv}^\mathrm{f}$ over all 
$\mathrm{f}\in F$, the \texttt{TBS} algorithm first constructs a minimum-cost tree 
$T$, where the cost of an edge connecting nodes 
$u$ and $v$ is given by 
$\Bar{\ell}_{uv}$. To ensure that $T$ is a valid topology, the greedy algorithm that generates $T$ traverses the list of pairs 
$u,v$ ordered by increasing 
$\Bar{\ell}_{uv}$, avoids creating vertices with a degree greater than 20, and ensures that at most one vertex has a degree exceeding 11.
The initial set 
$\mathcal{T}$ is subsequently defined as the collection of the 
$\kappa$ best neighboring topologies of 
$T$.

In summary, the \texttt{TBS} algorithm iteratively generates sets 
$\mathcal{T}$ of 
$\kappa$ topologies, and each of these topologies is evaluated using the function 
$f^{LB}$. The highest lower bound encountered during the algorithm is the value returned as the output by \texttt{TBS}.
For more details on this algorithm, for instance regarding the choice of parameter values such as the lengths of the tabu lists or the number 
$\kappa$ of elements in 
$\mathcal{T}$
, the reader is referred to references \cite{2} and \cite{Perreault2023}.

\subsection{Two new tabu search algorithms}\label{sec:5.2}

Although the \texttt{TBS} algorithm from the previous section uses tricks to bypass the complexity of determining the best neighbor topology of a given topology, each \texttt{TBS} iteration still requires a large amount of computation time: the number of neighboring topologies is in $O(n^3)$ (since $n{-}1$ edges can be removed and $O(n^2)$ edges can replace them), there are $O(n)$ possible master hubs, and the number of partitions of the master hub’s neighbors can reach several tens of thousands.
As a result, each \texttt{TBS} iteration takes a considerable amount of time, and only a few dozen iterations can be performed for a 50-node problem. Moreover, to speed up the evaluation of neighboring topologies, \texttt{TBS} uses an estimate of their value (i.e., the function $f^{Est})$, and a lower bound on the true value is computed for only a very small number of neighboring topologies. This is very risky, since it is not obvious that the neighboring topologies with the best estimated value are those that yield the best lower bound.

An avenue that was not investigated in \cite{2} as a means of circumventing the computational complexity associated with evaluating the value of a topology consists in avoiding the examination of all possible partitions of the master hub’s neighboring nodes. A key observation is that activating a greater number of beams simultaneously on an antenna leads to a reduction in signal quality. It therefore appears reasonable to group together nodes whose links to the master hub form similar angles.

For this reason, consider a master hub 
$r$ with $d$ neighboring nodes. We propose to first order these neighbors according to the angles formed by the links connecting them to 
$r$. Treating this ordered list as circular, we then construct an initial group by selecting 
$x$ consecutive nodes from the sequence, with 
$x$ varying between 
$\max\{0,d-10\}$ and 
$\min\{10,d\}$. The remaining 
$d-x$ neighbors of 
$r$ are assigned to a second block, thereby completing the partition of the master hub’s neighborhood. This procedure ensures that nodes with similar angular positions relative to the master hub are grouped together, which can help optimize signal quality considerations. 
In what follows, we write $P(T,r)$ for the set of these partitions associated with a master hub $r$ in a topology $T$.

\begin{figure}[!htb]
	\centering    \includegraphics[scale=0.75]{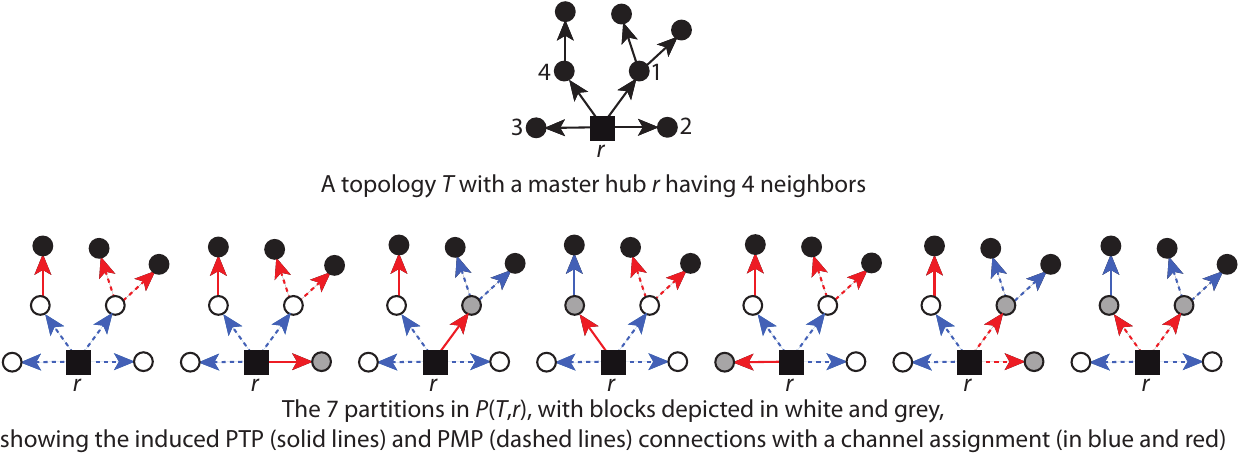}
\caption{Illustration of the considered partitions of the master hub's neighbors.}
	\label{fig:Partitions}
\end{figure}

It is straightforward to verify that the number of partitions in $P(T,r)$ is at most 55, since it equals $\frac{d(d-1)}{2}+1\leq 46$ when 
$d\leq 10$, whereas for 
$11\leq d\leq 20$ it is given by
$\frac{d(21-d)}{2}\leq 55$.
This is illustrated in Figure \ref{fig:Partitions}, where the two blocks are depicted in grey, and white, respectively. The sole partition excluded is the one in which nodes 1 and 3 are grouped together in one block, while nodes 2 and 4 are assigned to the other. Although in this illustrative example we consider 7 of the 8 possible partitions, the benefit of reducing the number of partitions to examine becomes far more pronounced as the number of neighbors of the master hub increases. For example, as discussed in Section \ref{sec:4}, when a master hub has 20 neighbors, there exist a total of 92,378 possible partitions; by contrast, the approach presented here considers only 10 partitions. Similarly, in the case of a master hub with 18 neighbors, our method yields 27 partitions, compared to the 68,068 possible partitions that would otherwise need to be evaluated. This demonstrates that the proposed strategy can substantially reduce computational complexity while still capturing the most relevant configurations

As with \texttt{TBS}, the solution space explored consists of all valid topologies that do not contain any edges from 
$E_{not}$. A neighboring topology is obtained by removing an edge and reconnecting the graph using another edge. Two tabu lists are used, $Tabu_r$ to forbid edge removals and $Tabu_a$ to forbid edge additions. Upon selecting a neighbor 
$T'$
of 
$T$ for the next iteration, the removed edge is added to $Tabu_a$, and the inserted edge is added to the list $Tabu_r$.

A lower bound on the value $f(T)$ of a topology $T$ is computed as described in Algorithm \texttt{LB1}. In words, only the nodes
$r\in R(T)$ may serve as master-hub candidates (see Section \ref{sec:tbs}). Then, the only partitions $\pi$ considered are those in 
$P(T,r)$. Also, given a partition of the master hub’s neighbors, an antenna configuration $\alpha$ is obtained with the technique described in Section \ref{sec:4}. Both channel assignments $\sigma$ are tested, and for each of them, a frequency assignment $\varphi$ is generated using the greedy algorithm also described in Section \ref{sec:4}. Thus, $O(T,r,\pi,\sigma,\varphi,\alpha)
$ is a lower bound on $f(T)$. 

\begin{algorithm}[!htb]
		\setstretch{0.98}
	\renewcommand\thealgorithm{}	
	\caption{\texttt{LB1}\\{\bf Input} : A valid topology $T$. \\
		{\bf Output} : A lower bound $LB$ on $f(T)$ as well as a master hub 
		$r_{LB}$, a partition 
		$\pi_{LB}$ of its neighbors, a channel assignment 
		$\sigma_{LB}$, a frequency assignment $\varphi_{LB}$, and an antenna configuraiton $\alpha_{LB}$.}
	\label{alg:LB1}
	\begin{algorithmic}[1]	
		\STATE $LB\gets -\infty$.
		\FOR{all $r\in R(T)$}
		\FOR{all $\pi\in P(T,r)$}
		\STATE Establish the antenna connfiguration $\alpha$ induced by $\pi$ (see Section \ref{sec:4}).
		\FOR{both possible channel assignements $\sigma$}
		\STATE Use the greedy algorithm to get a frequency assignement $\varphi$ (see Section \ref{sec:4}).
		\IF {$O(T,r,\pi,\sigma,\varphi,\alpha)>LB$}
		\STATE Set $LB{\gets} O(T,r,\pi,\sigma,\varphi,\alpha)$, $r_{LB}{\gets} r$, $\pi_{LB}{\gets} \pi$, $\sigma_{LB}{\gets} \sigma$, $\varphi_{LB}{\gets} \varphi$, and $\alpha_{LB}{\gets} \alpha$.
		\ENDIF
		\ENDFOR
		\ENDFOR
		\ENDFOR
	\end{algorithmic}
\end{algorithm}

The first tabu algorithm we present below is called \texttt{TABU1}. The initial topology is a minimum-cost tree, with the cost of an edge between nodes 
$u$ and $v$ defined as 
$\Bar{\ell}_{uv}$. The greedy algorithm used to construct 
$T$ ensures that the resulting tree is a valid topology (see Section \ref{sec:tbs}). The algorithm then functions like a standard tabu search, using \texttt{LB1} to determine the best neighbor. The algorithm stops when a time limit is reached.

\begin{algorithm}[!htb]
		\setstretch{0.98}
	\renewcommand\thealgorithm{}	
	\caption{\texttt{TABU1}\\{\bf Input} : A set of nodes. \\
		{\bf Output} : A valid topology $T_{best}$
		with a master hub 
		$r_{best}$, a partition 
		$\pi_{best}$ of its neighbors, a channel assignment 
		$\sigma_{best}$, a frequency assignment $\varphi_{best}$, and an antenna configuraiton $\alpha_{best}$.}
	\label{alg:T1}
	\begin{algorithmic}[1]	
			\STATE Generate an initial valid topology $T$ from a minimum-cost tree algorithm, with 
		$\Bar{\ell}_{uv}$ as the edge cost.
		\STATE Let $LB,r_{LB},\pi_{LB},\sigma_{LB},\varphi_{LB},\alpha_{LB}$ be the output of \texttt{LB1}$(T)$.
		\STATE Set $O_{best}{\gets} LB$, $T_{best}{\gets} T$, $r_{best}{\gets} r_{LB}$, $\pi_{best}{\gets} \pi_{LB}$, $\sigma_{best}{\gets} \sigma_{LB}$, $\varphi_{best}{\gets} \varphi_{LB}$, and $\alpha_{best}{\gets} \alpha_{LB}$.
		\STATE Set $Tabu_{a}\gets \emptyset$ and $Tabu_{r}\gets \emptyset$.
		
		\WHILE{the time limit is not reached}
			\STATE $O^*\gets -\infty$.
			\FOR{all edges $e$ in $T$ and not in $Tabu_{r}$}
				\STATE Let $C_1$ and $C_2$ be the two connected components resulting from the removal of $e$ from $T$.
				\FOR{all edges $e'\notin E_{not}\cup Tabu_{a}\cup\{e\}$ linking a vertex of $C_1$ to a vertex of $C_2$}
					\STATE Let $T'$ be the tree obtained from $T$ by replacing $e$ by $e'$.
					\IF{$T'$ is a valid topology}
						\STATE Let $LB,r_{LB},\pi_{LB},\sigma_{LB},\varphi_{LB},\alpha_{LB}$ be the output of \texttt{LB1}$(T')$
						\IF {$LB>O^*$}
							\STATE Set $T^*{\gets} T'$, $e_{r}{\gets}e$, $e_{a}{\gets}e'$ $O^*{\gets} LB$, $r^*{\gets} r_{LB}$, $\pi^*{\gets} \pi_{LB}$, $\sigma^*{\gets} \sigma_{LB}$, $\varphi^*{\gets} \varphi_{LB}$, and $\alpha^*{\gets} \alpha_{LB}$.
						\ENDIF
					\ENDIF
				\ENDFOR
			\ENDFOR
			\IF{$O^*>O_{best}$}
				\STATE Set $O_{best}{\gets} O^*$, $T_{best}{\gets} T^*$, $r_{best}{\gets} r^*$, $\pi_{best}{\gets} \pi^*$, $\sigma_{best}{\gets} \sigma^*$, $\varphi_{best}{\gets} \varphi^*$, and $\alpha_{best}{\gets} \alpha^*$.
			\ENDIF
			\STATE Add $e_{r}$ to $Tabu_{a}$ and $e_{a}$ to $Tabu_{r}$ and set $T{\gets} T^*$.
		\ENDWHILE
	\end{algorithmic}
\end{algorithm}

As will become evident in the following section, while \texttt{TABU1} allows for a greater number of iterations than the \texttt{TBS} algorithm for small- or medium-sized instances, each iteration remains computationally intensive and time-consuming. This observation motivates the design of a second algorithm, aimed at more efficiently identifying the best neighboring topology of a given topology 
$T$ without compromising solution quality.

The core idea underlying this second algorithm stems from the structural properties of the network. If a particular node is determined to be a good choice for the master hub in a topology 
$T$, it is highly likely that this node will also be a suitable candidate for the master hub in a neighboring topology 
$T'$ which is obtained from 
$T$ by replacing a single edge. This assumption is based on the fact that small local changes to the network, such as swapping one edge for another, generally do not drastically alter the centrality or relative importance of nodes within the network. Consequently, the master hub, being a critical and strategically positioned node, tends to remain a strong candidate across such minor modifications.

Similarly, the partition of the neighbors of the master hub is expected to remain largely stable between 
$T$ and its neighbor $T'$. While minor adjustments may be necessary to account for the addition or removal of a single edge, the overall structure of the partition typically does not change significantly. By exploiting this property, the second algorithm can focus its search on a smaller set of high-potential configurations, thereby significantly reducing the computational effort required to select the best neighbor.

The LB2 procedure described here below therefore assumes that the master hub 
$r$ and the partition 
$\pi$ of its neighbors are fixed, as well as the antenna configuration $\alpha$ (since it depends only on 
$r$ and $\pi$). A lower bound on 
$f(T)$ is thus obtained simply by testing the two channel assignments and using the greedy algorithm for the frequency assignment.

\begin{algorithm}[!htb]
		\setstretch{0.98}
	\renewcommand\thealgorithm{}	
	\caption{\texttt{LB2}\\{\bf Input} : A topology $T$, a master hub $r$, a partition $\pi$ of its neighbors, and an antenna configuration $\alpha$. \\
		{\bf Output} : A lower bound $LB$ on $f(T)$, a channel assignment 
		$\sigma_{LB}$, and a frequency assignment $\varphi_{LB}$}
	\label{alg:LB2}
	\begin{algorithmic}[1]	
		\STATE $LB\gets -\infty$.
		\FOR{both possible channel assignements $\sigma$}
		\STATE Use the greedy algorithm to get a frequency assignement $\varphi$ (see Section \ref{sec:4}).
		\IF {$O(T,r,\pi,\sigma,\varphi,\alpha)>LB$}
		\STATE Set $LB{\gets} O(T,r,\pi,\sigma,\varphi,\alpha)$, $\sigma_{LB}{\gets} \sigma$, and $\varphi_{LB}{\gets} \varphi$.
		\ENDIF
		\ENDFOR
	\end{algorithmic}
\end{algorithm}

Consider a topology 
$T$, a master hub 
$r$, a partition 
$\pi$ of its neighbors, and a neighboring topology 
$T'$ of $T$
obtained by replacing an edge 
$e$ of 
$T$ with an edge 
$e'$. If the removed edge 
$e$ connects 
$r$ to one of its neighbors 
$u$, then 
$u$ is removed from its block in the partition. Conversely, if the added edge 
$e'$ connects a vertex 
$v$ to 
$r$, then 
$v$ is inserted into the block containing the neighbor 
$w$ of 
$r$ whose incident angle with 
$r$ is most similar to that of 
$v$.
The resulting partition, denoted 
$\pi^T_{T'}$, is deemed invalid if any block contains more than 10 nodes, in which case the neighbor 
$T'$ of $T$ 
is discarded. The antenna configuration induced by the partition $\pi^T_{T'}$ in $T'$ is denoted $\alpha^T_{T'}$.

In \texttt{TABU2}, the neighbors 
$T'$ of $T$ are evaluated using the \texttt{LB2} procedure, where the master hub of 
$T'$
is identical to that of 
$T$, the partition of its neighbors is set to 
$\pi^T_{T'}$, and the antenna configuration to $\alpha^T_{T'}$.
To prevent the search from stagnating in local optima and to maintain diversity in the explored solutions, the choice of the master hub and the partition of its neighbors is periodically reconsidered every 
$\lambda$ iterations, where 
$\lambda$ is a user-defined parameter controlling the frequency of these restarts. Specifically, the algorithm performs a restart from the best topology obtained during the previous five iterations. At this point, both the master hub and the partition of its neighbors are determined using the \texttt{LB1} procedure, which exhaustively tests all possible choices to select the most promising configuration.
This mechanism ensures that \texttt{TABU2} not only exploits the current high-quality solutions but also systematically injects strategic diversification, allowing the algorithm to explore alternative configurations to  improve solution quality.

\begin{algorithm}[!htb]
		\setstretch{0.98}
	\renewcommand\thealgorithm{}	
	\caption{\texttt{TABU2}\\{\bf Input} : A set of nodes. \\
		{\bf Output} : A valid topology $T_{best}$
		with a master hub 
		$r_{best}$, a partition 
		$\pi_{best}$ of its neighbors, a channel assignment 
		$\sigma_{best}$, a frequency assignment $\varphi_{best}$, and an antenna configuraiton $\alpha_{best}$.}
	\label{alg:T2}
	\begin{algorithmic}[1]	
		\STATE Generate an initial valid topology $T$ from a minimum-cost tree algorithm, with 
		$\Bar{\ell}_{uv}$ as the edge cost.
		\STATE Let $LB,r_{LB},\pi_{LB},\sigma_{LB},\varphi_{LB},\alpha_{LB}$ be the output of \texttt{LB1}$(T)$.
		\STATE Set $O_{best}{\gets} LB$, $T_{best}{\gets} T$, $r_{best}{\gets} r_{LB}$, $\pi_{best}{\gets} \pi_{LB}$, $\sigma_{best}{\gets} \sigma_{LB}$, $\varphi_{best}{\gets} \varphi_{LB}$, and $\alpha_{best}{\gets} \alpha_{LB}$.
		\STATE Set $it\gets 0$, $r\gets r_{LB}$, $\pi\gets \pi_{LB}$, $\alpha\gets \alpha_{LB}$, $Tabu_{a}\gets \emptyset$ and $Tabu_{r}\gets \emptyset$.
		\WHILE{the time limit is not reached}
			\STATE $it\gets it+1.$
			\IF{$it$ is a multiple of $\lambda$}
				\STATE Let $T'$ be the topology that yielded the best lower bound over the last five iterations.
				\STATE Let $LB,r_{LB},\pi_{LB},\sigma_{LB},\varphi_{LB},\alpha_{LB}$ be the output of \texttt{LB1}$(T')$
				\STATE Set $T\gets T'$, $r\gets r_{LB}$, $\pi\gets \pi_{LB}$, and $\alpha\gets \alpha_{LB}$.
				\IF{$LB>O_{best}$}
					\STATE Set $O_{best}{\gets} LB$, $T_{best}{\gets} T$, $r_{best}{\gets} r_{LB}$, $\pi_{best}{\gets} \pi_{LB}$, $\sigma_{best}{\gets} \sigma_{LB}$, $\varphi_{best}{\gets} \varphi_{LB}$, and $\alpha_{best}{\gets} \alpha_{LB}$.
				\ENDIF
			\ELSE
				\STATE $O^*\gets -\infty$.
				\FOR{all edges $e$ in $T$ and not in $Tabu_{r}$}
					\STATE Let $C_1$ and $C_2$ be the two connected components resulting from the removal of $e$ from $T$.
					\FOR{all edges $e'\notin E_{not}\cup Tabu_{a}\cup\{e\}$ linking a vertex of $C_1$ to a vertex of $C_2$}
						\STATE Let $T'$ be the tree obtained from $T$ by replacing $e$ by $e'$.
						\IF{$T'$ is a valid topology and $\pi^T_{T'}$ is a valid partition}
							\STATE Let $LB,\sigma_{LB},\varphi_{LB}$ be the output of \texttt{LB2}$(T',r,\pi^T_{T'},\alpha^T_{T'})$
							\IF {$LB>O^*$}
								\STATE Set $T^*{\gets} T'$, $e_{r}{\gets}e$, $e_{a}{\gets}e'$ $O^*{\gets} LB$,  $\sigma^*{\gets} \sigma_{LB}$, and $\varphi^*{\gets} \varphi_{LB}$.
							\ENDIF
						\ENDIF
					\ENDFOR
				\ENDFOR
			\IF{$O^*>O_{best}$}
			\STATE Set $O_{best}{\gets} O^*$, $T_{best}{\gets} T^*$, $r_{best}{\gets} r$, $\pi_{best}{\gets} \pi(T\rightarrow T')$, $\sigma_{best}{\gets} \sigma^*$, $\varphi_{best}{\gets} \varphi^*$, and $\alpha_{best}{\gets} \alpha^*$.
			\ENDIF
			\STATE Add $e_{r}$ to $Tabu_{a}$ and $e_{a}$ to $Tabu_{r}$ and set $T{\gets} T^*$.
			\ENDIF
		\ENDWHILE
	\end{algorithmic}
\end{algorithm}

\section{Computational Experiments}
\label{sec:7}

To conduct a comprehensive evaluation of the two proposed algorithms, \texttt{TABU1} and \texttt{TABU2}, and to allow for a direct and rigorous comparison with the \texttt{TBS} algorithm described in \cite{2}, we utilized the same set of synthetic instances originally provided by our industrial partner. These instances represent networks of varying sizes, specifically 10, 15, 20, 30, and 50 nodes, covering a wide spectrum of problem complexities. For each network size, five instances were available, enabling statistically meaningful assessment of the algorithms’ performance; the exception is the largest size, 
$n=50$, for which only three instances were provided due to the significantly higher computational effort required to process these larger and more complex networks. This diverse set of instances thus allows us to evaluate not only the overall efficiency and effectiveness of the algorithms but also their scalability across networks of different sizes.

Based on a series of preliminary tests aimed at calibrating the performance of the algorithms, the parameters of \texttt{TABU1} and \texttt{TABU2} were established. Specifically, for a problem consisting of 
$n$ nodes, the parameter 
$\lambda$ is set equal to 
$n$, reflecting the desired frequency for periodically reconsidering the choice of the master hub and the partition of its neighbors. Furthermore, the tabu list 
$Tabu_r$ is configured to retain the last 
$\frac{\sqrt{n-1}}{2}$ elements that were inserted, while the tabu list 
$Tabu_a$
maintains the last 
$\sqrt{\tfrac{n(n-1)}{2}}$ elements. These settings were chosen to balance the need for exploration and diversification within the search process, ensuring that recently modified edges are temporarily prohibited from being reinserted or removed, thereby guiding the algorithm toward promising regions of the solution space while avoiding cycles or redundant moves.
	
For graphs containing 10 to 20 nodes, we imposed a time limit of 1 hour per instance. For instances with 30 nodes, this limit was extended to 2 hours, while for the largest instances, containing 50 nodes, the time limit was further increased to 5 hours in order to accommodate the substantially higher computational effort required for their processing.

All computational experiments were conducted on a workstation equipped with an Intel® Core™ i7-12700 processor (12th generation, 12 cores, 20 threads) and 64 GB of RAM, running a 64-bit AlmaLinux operating system. All algorithms were executed on CPU in single-threaded mode.

\begin{figure}[H]
\centering

\begin{subfigure}[b]{0.28\textwidth}
    \centering
    \includegraphics[width=\textwidth]{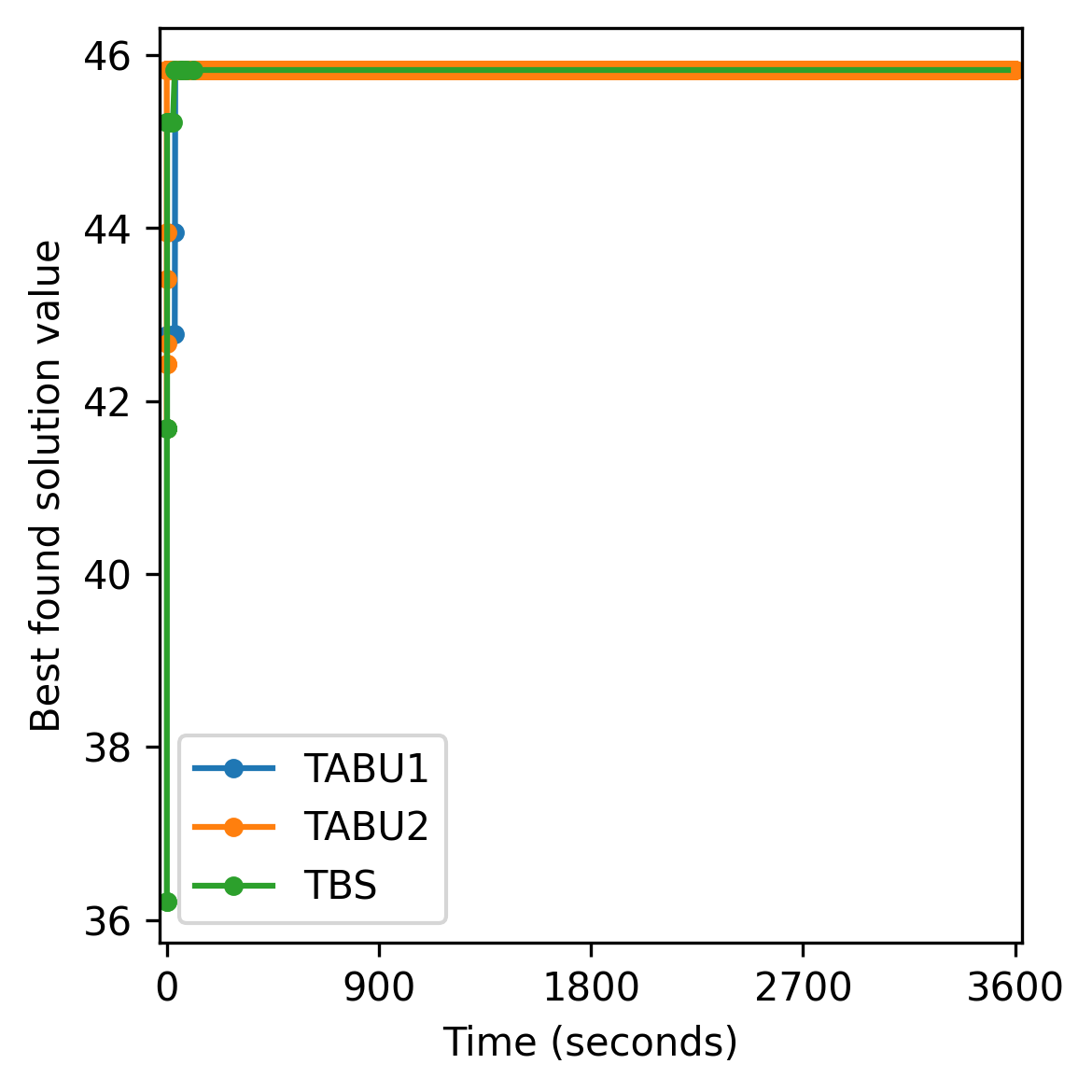}
    \caption{10 nodes}
\end{subfigure}
\hfill
\begin{subfigure}[b]{0.28\textwidth}
    \centering
    \includegraphics[width=\textwidth]{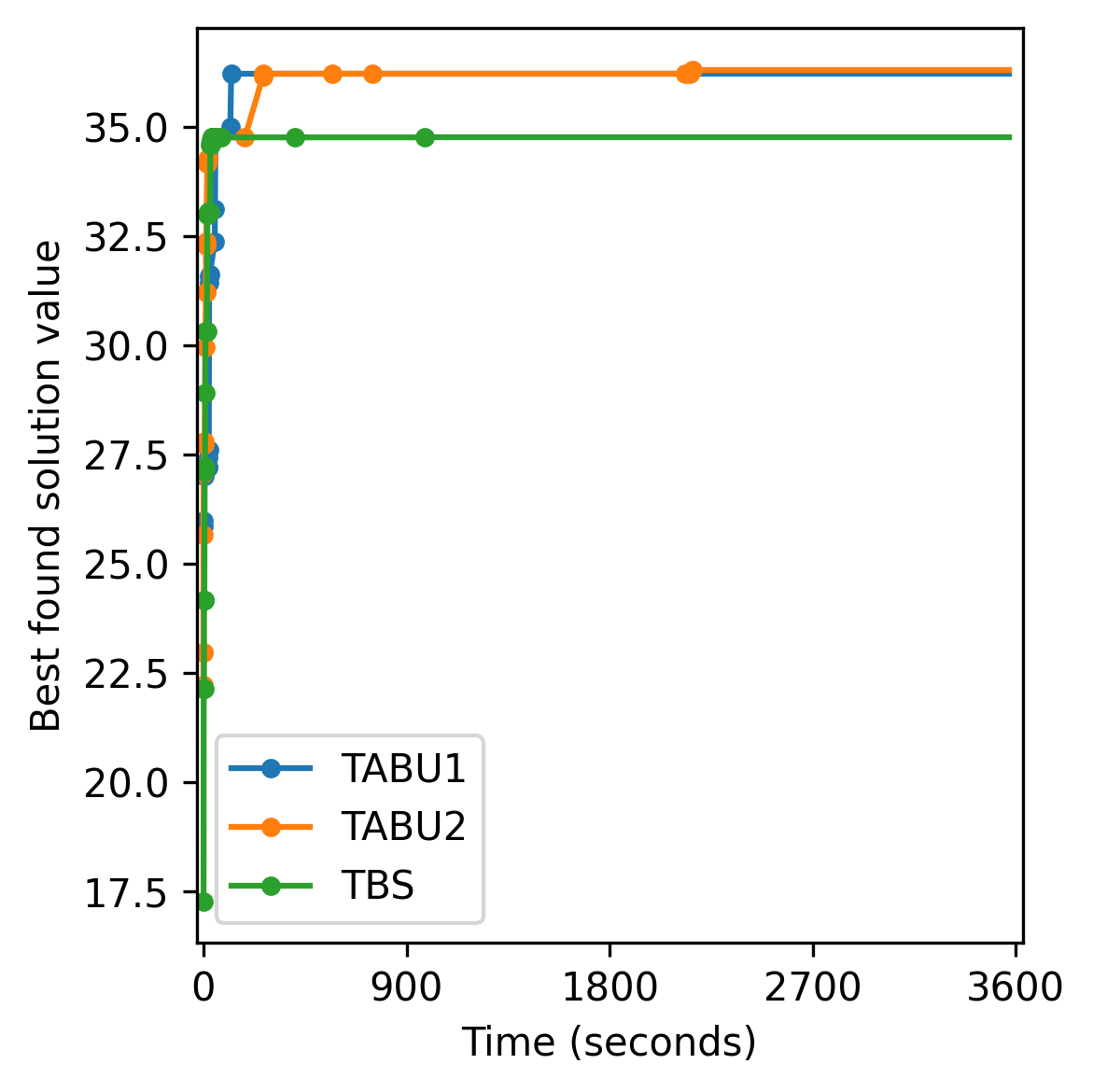}
    \caption{15 nodes}
\end{subfigure}
\hfill
\begin{subfigure}[b]{0.28\textwidth}
    \centering
    \includegraphics[width=\textwidth]{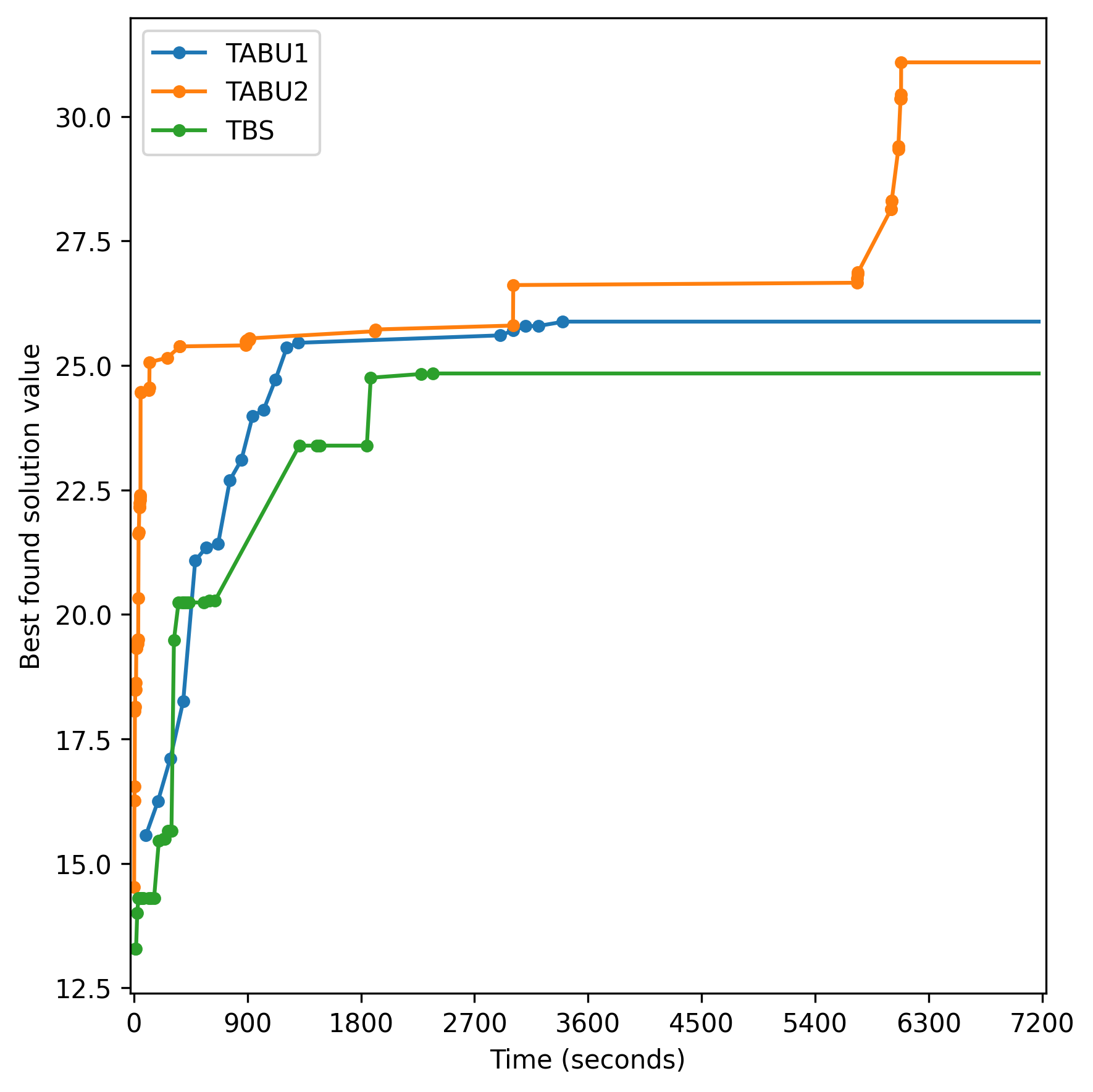}
    \caption{30 nodes}
\end{subfigure}

\caption{Evolution of the value of the best solution found as a function of time}
\label{fig:performance}
\end{figure}

We begin our analysis by presenting a selection of results in which we consider only a single instance for each of the network sizes 10, 15, and 30. This approach allows us to clearly illustrate the improvements in the number of iterations that can be achieved through the application of the \texttt{TABU1} and \texttt{TABU2} algorithms when compared to \texttt{TBS}, highlighting the efficiency and effectiveness of these methods. Table \ref{Table3instances} gives the values achieved as well as the number of iterations performed by each of the algorithms, while Figure \ref{fig:performance} shows the evolution of the value of the best solution found over time.

\begin{table}[!htb]
	\centering
	\caption{Performance comparison for three instances of sizes 10, 15, and 30 respectively }\label{tab:Table1}~\\[1ex]
	\label{Table3instances}\begin{tabular}{r|rrr|rrr} 
		\multicolumn{1}{c}{}&\multicolumn{3}{c}{solution value}&\multicolumn{3}{c}{iterations}\\
		\multicolumn{1}{c|}{$n$}&\texttt{TBS}&\texttt{TABU1}&\texttt{TABU2}&\texttt{TBS}&\texttt{TABU1}&\texttt{TABU2}\\
		\hline
		10&45.83&45.83&45.83&1,103&12,239&229,824\\
		15&34.77&36.23&36.32&770& 1,765&27,944\\
        30& 24.84 & 25.88 & 31.09 & 62& 76 &4,482\\
		\hline
	\end{tabular}
\end{table}

For the instance of size 10, its small size means that all the algorithms reach the same solution within a few seconds. A difference in performance becomes visible starting from 15 nodes. Upon examination of the results, we observe that \texttt{TABU2} executes a substantially larger number of iterations compared to both \texttt{TABU1} and \texttt{TBS}, which directly contributes to its ability to produce solutions of higher overall quality. While \texttt{TABU1} consistently outperforms \texttt{TBS}, the increase in iterations it achieves is insufficient to reach the level of solution quality delivered by \texttt{TABU2}. This highlights the effectiveness of the enhanced neighborhood exploration and strategic diversification mechanisms employed in \texttt{TABU2}, which allow it to more thoroughly explore the solution space and converge to superior configurations.

The results clearly indicate that the two proposed algorithms not only identify high-quality solutions, but also complete a larger number of iterations within the same execution window. For example, for the 10-node instance, \texttt{TABU2} executed approximately  230,000 iterations within a single hour, while \texttt{TBS} completed only around 1,100 iterations in the same time frame. For the 30-node instance, \texttt{TBS} manages to perform only 62 iterations in 2 hours, whereas \texttt{TABU2} performs 4,482. It should be noted, however, that while \texttt{TABU1}  performs more iterations than \texttt{TBS} for the instances of sizes 10 and 15, this is no longer the case for the instance of size 30. 


Large network instances render exhaustive evaluation of all master hub assignments and partitions infeasible due to combinatorial explosion. In such scenarios, minimizing per-iteration cost is essential, as it allows the algorithm to probe a much wider portion of the solution space. This broader and faster exploration is a key factor in maintaining solution quality and ensuring that the search remains effective even as complexity grows. 
This explains why \texttt{TABU1} and \texttt{TABU2} were able to reach better solutions than \texttt{TBS} for the instances of sizes 15 and 30.

Additional results are presented in Table \ref{tab:Table2} in order to provide a more comprehensive and insightful comparison of the performance of \texttt{TBS}, \texttt{TABU1} and \texttt{TABU2}. 
These results are based on problem instances taken from \cite{2}, covering sizes of 10, 15, 20, 30, and 50 nodes. For each size, five different instances were considered, except in the case of the 50-node instances, where only three instances were available. This selection of instances provides a wider and more representative range of problem sizes, allowing for a more thorough evaluation of how the methods perform across different levels of problem complexity. All algorithms were executed five separate times on each problem instance in order to account for the variability in their performance. For each instance, we report three key metrics regarding the solution values: the best solution value obtained across the five runs, the worst solution value, and the average solution value produced as output. In addition, for the number of iterations performed by the algorithms, we provide the minimum number of iterations recorded among the five executions, the maximum number observed, as well as the average number of iterations across the five runs. This detailed reporting allows for a comprehensive understanding of both the quality of the solutions found and the computational effort required by each algorithm.

\begin{table}[!htb]
\setlength{\tabcolsep}{4pt}\centering
\footnotesize
\caption{Performance comparison of \texttt{TBS}, \texttt{TABU1} and \texttt{TABU2} for instances with 10 to 50 nodes}
\label{tab:Table2}
\vspace{2mm}

\resizebox{1.0\textwidth}{!}{
\begin{tabular}{cc|rrr|rrr|rrr||rrr|rrr|rrr}
\multicolumn{2}{c}{} &
\multicolumn{9}{c||}{\textbf{solution value}} &
\multicolumn{9}{c}{\textbf{iterations}} \\
\multicolumn{2}{c}{} &
\multicolumn{3}{c}{TBS} &
\multicolumn{3}{c}{TABU1} &
\multicolumn{3}{c||}{TABU2} &
\multicolumn{3}{c}{TBS} &
\multicolumn{3}{c}{TABU1} &
\multicolumn{3}{c}{TABU2} \\
\multicolumn{1}{c}{$n$} &
\multicolumn{1}{c|}{instance} &
best & avg & worst &
best & avg & worst &
best & avg & worst &
min & avg & max &
min & avg & max &
min & avg & max \\
\hline


\multirow{6}{*}{10}																																						
&	1	&	45.83	&	45.83	&	45.83	&	45.83	&	45.83	&	45.83	&	45.83	&	45.83	&	45.83	&	714	&	1,170	&	1,461	&4,980	&	12,941	&	15,853	&	169,474	&	242,779	&	262,423	\\	
																																						
&	2	&	44.74	&	44.74	&44.74	&	44.74	&	44.74	&	44.74	&	44.74	&	44.74	&	44.74	&709	&	1,174	&	1,495	&	7,008	&	11,936	&	14,147	&	204,628	&	261,550	&	276,128	\\		
																																						
&	3	&	38.45	&	38.45	&38.45	&	38.45	&	38.45	&	38.45	&	38.45	&	38.45	&	38.45	&	710	&	1,130	&	1,449	&	14,867	&	15,808	&	18,003	&	229,619	&	310,752	&	331,457	\\	
																																						
&	4	&	47.07	&	47.07	&47.07	&	47.07	&	47.07	&	47.07	&	47.07	&	47.07	&	47.07	&	706	&	1,056	&	1,471	&	13,724	&	14,712	&	17,008	&	192,486	&	278,380	&	300,285	\\	
																																						
&	5	&	53.82	&	53.82	&53.82	&	53.82	&	53.82	&	53.82	&	53.82	&	53.82	&	53.82	&	704	&	932	&	1,104	&	12,400	&	13,622	&	17,282&	171,341	&	251,052	&	272,118	\\		
																																						
\cline{2-20}&average&	45.98	&	45.98	&	45.98	&45.98	&45.98	&45.98	&45.98	&45.98	&45.98	&708	&	1,092	&	1,396	&10,596	&13,804	&16,459	&193,510	&268,903	&288,482\\

\hline\hline																																						
																																						

\multirow{5}{*}{15}																																						
&	1	&	34.77	&	34.77	&	34.77	&	36.23	&	36.23	&	36.23	&	36.32	&	36.32	&	36.32	&	451	&	636	&	772	&	1,640	&	1,784	&	2,163	&	23,770	&	28,866	&	30,448	\\
																																						
&	2	&	35.26	&	34.99	&	34.82	&	35.26	&	35.26	&	35.26	&	36.25	&	36.25	&	36.25	&	117	&	589	&	800	&	931	&	1,206	&	1,336&	20,303	&	27,288	&	29,150	\\	
																																						
&	3	&	34.26	&	34.24	&	34.17&	34.26	&	34.24	&	34.14&	35.27	&	35.27	&	35.27&	445	&	653	&	793	&	742	&	1,484	&	1,773&	33,616	&	41,236	&	43,756	\\				
																																						
&	4	&	34.11	&	33.47	&	31.96&	33.08	&	33.08	&	33.08&	34.11	&	34.11	&	34.11&	460	&	649	&	787	&	1,461	&	1,539	&	1,654&	33,335	&	42,925	&	45,725	\\				
																																						
&	5	&	33.05	&	32.13	&	29.73&	33.05	&	33.05	&	33.05&	33.05	&	33.05	&	33.05&	272	&	633	&	835&	1,415	&	1,616	&	2,184&	31,934	&	42,352	&	45,380	\\					
																																						
\cline{2-20}&average&	34.29	&	33.92	&	33.09	&34.38	&34.37	&34.35	&35.00	&35.00	&35.00	&349	&	632	&	797	&	1,238&	1,526&	1,822&	28,592&	36,533&	38,892\\

\hline\hline

																																						
\multirow{5}{*}{20}																																						
&	1	&	33.96	&	33.96	&	33.96&	33.35	&	33.35	&	33.35&	34.53	&	34.53	&	34.53&	164	&	166	&	166	&	241	&	279	&	378&	8,799	&	12,413	&	13,377	\\				
																																						
&	2	&	30.96	&	30.88	&	30.56&	31.75	&	31.75	&	31.75&	31.66	&	31.66	&	31.66&	230	&	261	&	376	&	245	&	285	&	381&	7,449	&	10,226	&	10,973	\\				
																																						
&	3	&	32.29	&	31.14	&	30.29&	31.69	&	31.69	&	31.69&	33.66	&	33.66	&	33.66&	152	&	161	&	180	&	220	&	256	&	352&	9,465	&	13,091	&	14,063	\\				
																																						
&	4	&	32.46	&	31.36	&	30.49&	32.46	&	32.46	&	32.46&	32.46	&	32.46	&	32.46&	257	&	296	&	367	&	249	&	312	&	441&	9,947	&	13,824	&	14,850	\\				
																																						
&	5	&	31.79	&	31.56	&	30.88&	31.99	&	31.71	&	31.62&	31.99	&	31.99	&	31.99&	173	&	193	&	226	&	218	&	262	&	374&	9,184	&	12,519	&	13,394	\\				
																																						
\cline{2-20}&average&	32.29	&	32.18	&	31.04	&32.24	&32.19	&32.17	&32.86	&32.86	&32.86	&195	&	215	&	263	&	235&	279&	385&8,969&	12,415&	13,331\\																
																																						
\hline\hline																																						
																																						

\multirow{5}{*}{30}																																						
&	1	&	23.21	&	23.21	&	23.21&	25.8	&	25.53	&	25.16&	26.67	&	26.26	&	24.63&	38	&	47	&	49	&	49	&	65	&	77&	2,924	&	4,215	&	4,561	\\				
																																						
&	2	&	20.24	&	17.85	&	16.55&	27.58	&	27.58	&	27.58&	25.26	&	25.23	&	25.11&	70	&	136	&	153	&	53	&	67	&	79&	2,682	&	3,868	&	4,182	\\				
																																						
&	3	&	21.83	&	20.07	&	15.99&	22.27	&	21.71	&	20.88&	27.63	&	27.63	&	27.63&	57	&	83	&	89	&	51	&	65	&	77&	2,820	&	3,869	&	4,146	\\				
																																						
&	4	&	24.84	&	23.01	&	19.02&	26.69	&	26.2	&	25.88&	31.09	&	30.2	&	26.62&	62	&	64	&	69	&	52	&	68	&	79&	3,098	&	4,350	&	4,674	\\				
																																						
&	5	&	18.51	&	17.93	&	16.24&	24.75	&	24.75	&	24.75&	24.85	&	24.84	&	24.78&	65	&	109	&	122	&	46	&	61	&	71&	2,498	&	3,730	&	4,046	\\				
\cline{2-20}&average&	21.73	&	20.41	&	18.4	&25.42	&25.16	&24.85	&27.10	&26.83	&25.75	&58	&	88	&	96&	50&65&	77&	2,804&	4,006&	4,322\\																	
																																						
\hline\hline																																						
																																						

\multirow{4}{*}{50}																																						
&	1	&	9.27	&	9.27	&	9.27&	11.77	&	11.39	&	11.24	&	16.98	&	16.98	&	16.98&	62	&	76	&	86	&	7	&	8	&	10&	843	&	861	&	866	\\			
																																						
&	2	&	7.72	&	6.89	&	5.97&	10.19	&	8.38	&	7.86&	18.81	&	18.81	&	18.81&	44	&	63	&	76	&	6	&	7	&	9&	1,069	&	1,084	&	1,098	\\				
																																						
&	3	&	10.35	&	8.61	&	7.73&	11.00	&	10.64	&	10.49&	18.47	&	18.47	&	18.47&	76	&	93	&	141	&	6	&	7	&	10&	929	&	941	&	947	\\				
																																						
\cline{2-20}&average&	9.11	&	8.26	&	7.66	&10.99	&10.14	&9.86	&18.09	&18.09	&18.09	&61	&	77	&	101	&	6&7&	10&	947&	962&	970\\																
																																						
\hline

\end{tabular}
} 
\end{table}

We can observe in Table \ref{tab:Table2} that for the larger instances, the improvement in solution quality becomes particularly striking, reaching increases of about 150\% in some cases. To illustrate this, consider the second instance with 50 nodes: whereas \texttt{TBS} manages to obtain a solution with a value of only 7.72, the \texttt{TABU2} method identifies a significantly superior solution with a value of 18.81. This example clearly highlights the substantial advantage provided by the optimized approach when dealing with high-complexity problem sizes.

It therefore becomes apparent from the data presented in Table \ref{tab:Table2} that the performance gap between \texttt{TBS} and \texttt{TABU2} becomes increasingly apparent as the size of the instances grows. This observation directly aligns with one of the primary goals of our study, which was to evaluate how well each approach scales when confronted with larger and more demanding network configurations. As the network size expands, the inherent complexity of the problem rises dramatically, not only does the number of possible configurations increase, but the computational effort required to evaluate and compare candidate solutions also becomes substantially heavier. This escalation affects both the number of iterations that can be executed within a fixed time limit and the ability of the algorithm to converge toward high-quality solutions.

Despite these challenging conditions, \texttt{TABU2} consistently demonstrated a strong capacity to sustain a high iteration rate. Within the same execution time budget, it managed to evaluate far more configurations than \texttt{TBS}, allowing it to navigate the solution space with much greater depth and breadth. As a direct consequence of this expanded exploration capability, our approach regularly achieved superior objective values, even on the most complex instances.

With regard to \texttt{TABU1}, it is observed that this algorithm occasionally produces better solutions than \texttt{TABU2}; specifically, this occurs for two instances with 20 nodes and one instance with 30 nodes. This can be explained by the fact that although \texttt{TABU1} runs fewer iterations than \texttt{TABU2}, it allows the master hub and the partitioning of its neighbors to be modified at each iteration, whereas \texttt{TABU2} only allows such modifications occasionally.
However, when considering all instances, \texttt{TABU2} generally yields better solutions than \texttt{TABU1}. 

Compared to \texttt{TBS}, \texttt{TABU1} tends to perform better in most cases, although there are exceptions: it performs worse than \texttt{TBS} on one instance with 15 nodes and on two instances with 20 nodes. Interestingly, even though \texttt{TABU1} carries out fewer iterations than \texttt{TBS} on the larger instances with 30 and 50 nodes, it is still able to reach solutions of higher quality, demonstrating that the efficiency of an algorithm in terms of solution quality is not solely determined by the number of iterations performed. This can be explained by the fact that \texttt{TBS} evaluates neighboring topologies by estimating their value using the function 
$f^{Est}$
, and only computes a lower bound on the optimal value for a very limited number of neighboring topologies. In contrast, \texttt{TABU1} assigns frequencies to each node of a neighboring topology in order to obtain a lower bound on the best achievable value. Since computing a lower bound is more computationally expensive than the estimation (which assumes no interference and therefore does not require assigning frequencies), \texttt{TABU1} evaluates fewer neighboring topologies than \texttt{TBS}, but the value of each evaluated topology is not just a simple estimate, it is a value that is actually achievable. In addition, as discussed in Section \ref{sec:5.2}, relying on 
$f^{Est}$ is highly risky, since there is no clear guarantee that the neighboring topologies with the best estimated values are those that produce the strongest lower bound.

Overall, these findings highlight the efficiency of our Tabu Search variants when dealing with large-scale problems. They confirm that the proposed methods remain effective and reliable in situations where more traditional or exhaustive strategies tend to plateau or deteriorate in performance, thereby validating their suitability for tackling increasingly large network design problems.\\


\section{Conclusions and future work}
\label{sec:8}

In this study, we proposed a thoroughly optimized and methodically refined approach for the design of tactical network topologies, harnessing the advanced capabilities of a Tabu Search algorithm. To manage the inherent complexity of the network design problem, we decomposed it into a sequence of clearly defined and easily manageable sub-tasks. This decomposition allowed us to systematically address each critical aspect of the network, including the selection of a master hub, the strategic partitioning of its neighboring nodes, the careful assignment of channels and frequencies, and the precise configuration of antennas. By approaching the problem in this structured and stepwise manner, we ensured that all elements of the network were accounted for in a consistent, coherent, and well-organized framework, thereby facilitating more effective exploration of high-quality network configurations.

A key innovation in our approach was the use of geometric functions to partition the set of successors of the master hub. This strategic partitioning allowed us to drastically reduce the otherwise exponential number of potential partitions, effectively limiting the number of candidate configurations that needed to be evaluated during each iteration of the search. Streamlining this part of the algorithm enabled the Tabu Search to explore neighborhoods more efficiently, allowing more iterations within the same computational budget. Compared to the more restricted iteration framework used in the study reported in \cite{2}, our method provides a significantly broader and deeper exploration of the solution space, which in turn enhances the likelihood of identifying high-quality network configurations.


For future research directions, building upon the approach explored in the study by \cite{liu2023machine}, the performance of the proposed algorithms could be substantially enhanced through the integration of advanced Machine Learning techniques aimed at optimizing critical decision points. One promising avenue is the use of Graph Neural Networks (GNNs) for node prediction, which would enable the identification of the most promising candidates for selection as the master hub. By effectively narrowing down the pool of potential hub nodes, this strategy would considerably reduce the number of configurations that the algorithm must evaluate, thereby accelerating the overall search process.

In addition to master hub selection, Machine Learning methods could play a pivotal role in the partitioning of the master hub’s successors. In this context, a GNN could be trained to predict near-optimal partitions, streamlining the assignment of channels and frequencies while minimizing potential interference. Such predictive guidance would allow the search process to focus on configurations that are more likely to yield high-quality solutions, reducing wasted computational effort on less promising alternatives.

Moreover, learning-based models could be employed during neighborhood generation to anticipate which edges should be added or removed, effectively guiding the algorithm toward the most promising neighbors in the solution space. By leveraging these predictive insights, the Tabu Search would be able to concentrate its computational resources on evaluating topologies with the highest potential, resulting in a more efficient and targeted exploration of the search space. Overall, the integration of Machine Learning techniques holds significant promise for improving both the speed and effectiveness of \texttt{TABU2}, particularly in complex, large-scale network design problems where exhaustive exploration is computationally prohibitive.

In the present study, all constraints and parameter values were established based on the specific requirements and specifications provided by our industrial partner. While this approach ensured that the network design was aligned with practical, real-world needs, an important avenue for future research lies in critically examining the sensitivity of these fixed parameters and constraints. Understanding how variations in key parameters influence the performance of the network and the behavior of the optimization algorithm is essential for assessing the generalizability of the proposed approach.

For example, one pertinent question concerns the number of beams per antenna. In the current study, antennas were assumed to have 24 beams, consistent with the hardware specifications provided. However, it is natural to ask how the overall network performance would be affected if antennas with fewer beams, such as 6 or 12, were considered instead. By systematically varying this parameter and analyzing the resulting impact on network efficiency, solution quality, and the computational characteristics of the algorithm, it would be possible to gain deeper insight into the interplay between hardware capabilities and optimization outcomes.

Conducting such a sensitivity analysis would provide valuable guidance for adapting the proposed methodology to a wider range of hardware configurations and operational contexts. It would also help quantify the robustness of the algorithm under different assumptions and inform practical decision-making when balancing design choices, performance targets, and computational resources. Ultimately, this line of investigation would strengthen the applicability of the approach across diverse scenarios and enhance confidence in its use for real-world tactical network design problems.\\


\noindent\textbf{Declaration of Competing Interest}\\
The authors declare that they have no competing interests. No funding was received from any organization that could be perceived as influencing the research presented in this paper.\\

\noindent\textbf{Credit authorship contribution statement}\\
Wissem Ahmed Zaid is the leading contributor of this work, responsible for the algorithmic design (50\%) and the entirety of the experimental evaluation (100\%).
Alain Hertz contributed to the algorithmic design (50\%) in his role as supervisor.
Both authors contributed equally to the writing of the paper.\\

\noindent\textbf{Data availability}\\
The data used for testing in this study are the same as those generated by \cite{2}.\\

\noindent\textbf{Acknowledgements}\\
We would like to sincerely thank Oussama Siwane for his invaluable technical assistance in translating the method presented in \cite{2} from Python to C++. His expertise and careful work were instrumental in ensuring that the original algorithm was accurately and effectively implemented in the new programming environment. We also wish to extend our heartfelt gratitude to Patrick Munroe for his guidance and support in optimizing the resulting implementation. His contributions were crucial in enhancing the efficiency and performance of the code, enabling it to run smoothly and reliably, which was essential for the successful execution of our experiments.


\bibliographystyle{acm} 
\bibliography{REF}

\newpage
\section{Appendix}\label{appendix}

Let 
$V$ denote the set of nodes in a network design instance, with each node 
$v\in V$ having known coordinates 
$(x_v,y_v)$. Let $T$ be a tree topology on $V$, and let $\varphi$ be a frequency assignment on $T$.  Let $uv$ be an edge in $T$, and suppose frequency $f$ is used on $uv$.  Suppose also that the antenna configuration at each node is fixed. In order to compute the direct throughput $TP_{uv}$ on $uv$, we first compute the antenna gain $g_{uv}^f$ at node $u$ in
the direction of node $v$, with angle $\phi$ and with a set $B$ of active beams. This is given (in decibels) by 
 \begin{equation*}
	g_{uv}^{f} = 10 \, \log_{10} \sum_{b \in B} \textrm{exp}_{10} \left(\left.g_{uv,b}^{f}\right/10\right),
\end{equation*}
\noindent with
\begin{equation*}
	\begin{aligned}\label{eq:guvbf}
		g_{uv,b}^{f} ={} & g_{\textrm{max}}(|B|, f) 
		& - 3 \, \frac{\log_{10} \left(\cos \left(\Delta_{\phi} (\phi + b \frac{2 \pi}{24},x_u,y_u,x_v,y_v)\right) \right)_{+}}{\log_{10} \left(\cos \left(\Delta_{3dB}(f) / 2\right)\right)_{+}},
	\end{aligned}
\end{equation*}
where 
\begin{itemize}\itemsep=0pt
	\item $g_{\textrm{max}}(|B|, f)$ is the gain in  decibels in the maximal direction of each beam, \item $\Delta_{\phi} (\phi + b \frac{2 \pi}{24},x_u,y_u,x_v,y_v)$ is the angle deviation between this maximal direction and the node $v$ relative to node $u$, 
	\item $\Delta_{3dB}(f)$ is the 3 dB beam width (the width of the beam at which there is a 3 dB loss in signal) and 
	\item $(\cdot)_{+} = \max \{\cdot, 0\}$.
\end{itemize} We use
$$
g_{\textrm{max}}(|B_a|, f) = \left\{ \begin{array}{ll}
	13 - 10 \log_{10}(|B|), & \textrm{if } \textrm{channel}(f) = 3+ \\
	15 - 10 \log_{10}(|B|), & \textrm{if } \textrm{channel}(f) = 4,
\end{array} \right.
$$
and
$$
\Delta_{3dB}(f) = \left\{ \begin{array}{ll}
	60^{o}, & \textrm{if } \textrm{channel}(f) = 3+ \\
	50^{o}, & \textrm{if } \textrm{channel}(f) = 4.
\end{array} \right.
$$

The gain $g_{vu}^f$ in the opposite direction is computed in a similar way. Given the antenna gains in both directions and the path loss $p_{uv}^f$ between the two nodes, we can then compute the signal strength (in decibel-milliwatts)
\begin{equation*}\label{eq:suvf}
	s_{uv}^f = 30 + g_{uv}^f + g_{vu}^f - p_{uv}^f.
\end{equation*}

These signal strengths are computed between every pair of antennas that use the same frequency and not only the pairs that are actually connected in the network. The extra signal from the unconnected pairs that use the same frequency creates interference on the actual connections. The resulting interference between connected nodes $u$ and $v$ on frequency $f$ is given (in milliwatts) by
\begin{equation*}
	i_{uv}^f = \sum_{\substack{w\neq u,v \textrm{ with an antenna}\\ \textrm{that uses frequency } f }} \left( \textrm{exp}_{10} \left(\frac{s_{uw}^f}{10}\right) + \textrm{exp}_{10} \left(\frac{s_{wv}^f}{10}\right) \right).
\end{equation*}

With the interference, we can compute the Signal-to-Noise Ratio (SINR) of the connection between nodes $u$ and $v$ (in decibel-milliwatts) with
\begin{equation*}\label{eq:Suvf}
	S_{uv}^f = s_{uv}^f - m_{uv}^f - 10 \, \log_{10} \left( i_{uv}^f + \textrm{exp}_{10} \left(\frac{N\!P}{10}\right) \right),
\end{equation*}
where $m_{uv}^f$ is the fade margin and $N\!P$ is the receiver antenna's noise power. Assuming a 20 MHz bandwidth and a 10 dB noise figure, we use
$$
N\!P = -174 + 10 \, \log_{10} \left(20 \cdot 10^6\right) + 10.
$$

\noindent The direct throughput $TP_{uv}$ is then finally taken from the following table:\\

\centering	\begin{tabular}{  c  c  } 
		\toprule
		SINR $S$ & Throughput \\ 
		\midrule
		$S < 2$ & 0 \\ 
		$2 \leq S < 5$ & 6.5 \\ 
		$5 \leq S < 9$ & 13 \\ 
		$9 \leq S < 11$ & 19.5 \\ 
		$11 \leq S < 15$ & 26 \\ 
		$15 \leq S < 18$ & 39 \\ 
		$18 \leq S < 20$ & 52 \\ 
		$20 \leq S < 25$ & 58.5 \\ 
		$25 \leq S < 29$ & 65 \\ 
		$29 \leq S$ & 78 \\ 
		\bottomrule
	\end{tabular}


\end{document}